\documentclass[showpacs,twocolumn,amsmath,amssymb,pra]{revtex4-1}

\usepackage{amsmath} 

    \usepackage{graphicx}
    \usepackage{subfigure}
    \usepackage{color}
    \usepackage{dcolumn}
    \usepackage{bm}
    \usepackage{xspace}
\usepackage[colorlinks=true, urlcolor=blue, pdfborder={0 0 0}]{hyperref}

    \newcommand{\schr}{Schr\"o\-din\-ger\xspace}

    \newcommand{\IgnoreThis}[1]{#1}

\begin{document}

\title{Equivalence between free quantum particles and those in
  harmonic potentials\\and its application to instantaneous changes}

\author{Ole Steuernagel}

\affiliation{School of Physics, Astronomy and Mathematics, University of
Hertfordshire, Hatfield, AL10 9AB, UK }
\email{O.Steuernagel@herts.ac.uk}

\date{\today}

\begin{abstract}
  In quantum physics the free particle and the harmonically trapped
  particle are arguably the most important systems a physicist
  needs to know about. It is little known that, mathematically, they
  are one and the same. This knowledge helps us to understand either
  from the viewpoint of the other. Here we show that all general
  time-dependent solutions of the free-particle \schr equation can be
  mapped to solutions of the \schr equation for harmonic potentials,
  both the trapping oscillator and the inverted `oscillator'. This map
  is fully invertible and therefore induces an isomorphism between
  both types of system, they are equivalent. A composition of the map
  and its inverse allows us to map from one harmonic oscillator to
  another with a different spring constant and different center
  position. The map is independent of the state of the system,
  consisting only of a coordinate transformation and multiplication by
  a form factor, and can be chosen such that the state is identical in
  both systems at one point in time. This transition point in time can
  be chosen freely, the wave function of the particle evolving in time
  in one system before the transition point can therefore be linked up
  smoothly with the wave function for the other system and its future
  evolution after the transition point. Such a cut-and-paste procedure
  allows us to describe the instantaneous changes of the environment a
  particle finds itself in. Transitions from free to trapped systems,
  between harmonic traps of different spring constants or center
  positions, or, from harmonic binding to repulsive harmonic
  potentials are straightforwardly modelled. This includes some
  time-dependent harmonic potentials. The mappings introduced here
  are computationally more efficient than either state-projection or
  harmonic oscillator propagator techniques conventionally employed
  when describing instantaneous (non-adiabatic) changes of a quantum
  particle's environment.
\end{abstract}


\maketitle

\IgnoreThis{\section{Introduction and Motivation}}

A quantum particle confined by a harmonic potential features a
discrete energy spectrum; a free particle's energy can have any
positive value. At first, it might sound implausible that these two
cases are fully equivalent, yet, simple mathematics accessible at the
undergraduate level suffices to prove this fact.

This equivalence provides a formal connection between the free and the
trapped case helping us to understand features of one system from the
point of view of the other~\cite{Ole_AJP05}, or why they share
features~\cite{Ole_PRL13}. It can also help with calculations. If
analytical solutions for a trapped system are needed one can switch to
the free particle picture since the propagators are easier to
integrate and therefore closed form solutions may be determined which
are otherwise inaccessible~\cite{Ole_arXiv1109.1818}; then one can map
back to the trapped system. For numerical calculations the trapped
particle picture can offer advantages because the confinement of the
particle allows us to circumvent grid adaptation, often necessary for
modelling of free particles spreading without bound. Here, particular
attention is paid to the application of the equivalence for the
modelling of instantaneous transitions. Three types of transitions can
easily be described: instantaneously setting a particle free or capturing a
free particle, instantaneous change of the spring constant or
stiffness of a harmonic trap, including instantaneous switch-over to
an inverted `oscillator', and instantaneous displacements of a trap.

That the cases of free and harmonically trapped quantum particle are
connected seems to have been observed first in the field of optics,
rather than in the field of mathematical physics or quantum
mechanics. For instance Yariv's textbook~\cite{Yariv.book_67} from
1967 introduces the well known freely propagating modes of laser beams
and points out that they are ``of the same form'' as the eigenstates
of the harmonic oscillator. In mathematical physics their full
equivalence appears to have been established first by Niederer in
1972~\cite{Niederer_HPA72} using group theoretical arguments and by
Takagi in 1990~\cite{Takagi_PIT90} using coordinate
transformations. Takagi's work additionally emphasises that a
continuous change of spring constant of the harmonic binding potential
over time can be modelled. An associated constant of motion for
time-varying harmonic potentials was investigated by Lewis in
1967~\cite{Lewis_PRL67}. In theoretical optics the equivalence between
free and harmonically trapped particles was investigated by Nienhuis
\emph{et al.} in 1993 using operator methods~\cite{Nienhuis_PRA93}.
The connection with the Gouy-phase of optics, the phase shift of $\pi$
(or multiples thereof) a beams suffers when going through a focus, was
made explicit by Steuernagel in 2005~\cite{Ole_AJP05}. Barton's
comprehensive 1986 article on the inverted
`oscillator'~\cite{Barton_AnnP86} does not mention the equivalence
although it also applies to such a system (which disperses a particle
with a negative spring constant, i.e., a repulsive Hookean force). A
connection to the inverted oscillator has been made in 2006 by Yuce
and others~\cite{Yuce_PhysScrpt06}. Recently, interesting features
of freely moving waves have been analyzed using the equivalence~\cite{Andrzejewski_13arXiv1310.2799}.

Here, we give one possible classical mechanics motivation for the
equivalence in section~\ref{sec_Classical}. The map for instantaneous
harmonic trapping of a free particle and its inverse is presented in
section~\ref{sec_Maps}, followed by its application to the modelling
of instantaneous changes in the environment in
section~\ref{sec_DetailsMap}. It is one of the main motivations for
this article to show that the equivalence can be used to describe
instantaneous (non-adiabatic) changes in simple terms; it seems that this
application has not been reported before although it is
interesting~\cite{Kiss_PRA94,MoyaCessa_PLA03}. A composition of
different maps to model mappings between harmonic potentials of
different spring constants is introduced in
section~\ref{sec_compose}. This is followed by mappings between
laterally shifted traps of different spring constants in
section~\ref{sec_LatShift}. Mappings to inverted `oscillators' in
section~\ref{sec_Inv_Osc} are followed by mappings to freely falling
(uniformly accelerated) particles in
Section~\ref{sec_freeFall}. Section~\ref{sec_generalized} shows that
the approach described here cannot be extended to other systems such
as particles in anharmonic potentials; the following
section~\ref{sec_timeDependent} does establish the class of
time-dependent harmonic potentials free particles can be mapped to. We
conclude with a summary of the usefulness of our results in
section~\ref{sec_concl}.

\section{Classical Case}\label{sec_Classical}

To motivate the quantum case let us briefly consider the 1-dimensional
classical harmonic oscillator with coordinate~$\xi$, described by the
Hamiltonian
\begin{eqnarray}
  H = \frac{p^2}{2M} + \frac{k}{2} \xi^2 = 
\frac{1}{2M}(p^2 + M^2 \omega^2 \xi^2) \; ,
\label{eq_ClassHamiltonian}
\end{eqnarray}
where $M$ is the mass of the particle, $k$ the spring constant and
~$\omega = \sqrt{k/M}$ its resonance frequency. In suitable
coordinates its phase space trajectories are circles, this suggests
the coordinate transformation (paraphrasing
Goldstein~\cite{Goldstein.book_81})
\begin{eqnarray}
\xi(X,P) & = &   \sqrt{\frac{2 P}{ M \omega}} \cos(X) \\
  \mbox{and } \quad   p(X,P) & = & - \sqrt{2 P M \omega} \sin(X) \; .
\label{eq_ClassCoordTrafo}
\end{eqnarray}
These are known to form `canonical transformations', which implies
that they map $H$ to the equivalent
Hamiltonian~\cite{Goldstein.book_81}
\begin{eqnarray}
  H' = \omega P (\sin(X)^2+\cos(X)^2) = \omega P \; ,
\label{eq_ClassHamiltonian2}
\end{eqnarray}
which is independent of the coordinate $X$, that is, the generalized
momentum~$P$ is conserved. Using the Hamiltonian equation of motion
$\frac{d}{dt}{X} = \partial H' / \partial P = \omega $ yields
\begin{eqnarray}
  X(t)=\omega t + \phi_0 \; .
\label{eq_ClassSolution}
\end{eqnarray}
Upon insertion into Eq.~(\ref{eq_ClassCoordTrafo}),
Eq.~(\ref{eq_ClassSolution}) not only yields the correct solution but
structurally it resembles the motion of a free particle $x(t) = v t
+x_0$.

To extract time via the phase angle of a harmonic oscillator one can
form the arctangent of its circular motion, it is therefore perhaps
not surprising that the correct guess for the coordinate mapping for
time in the quantum case involves tangent and arctangent functions,
compare Eqs.~(\ref{eq_t_map}) and~(\ref{eq_tau_inv_map}) below.

\section{Map from Free to Trapped Case}\label{sec_Maps}

The \schr equation for a free particle of mass~$M$ in one spatial
dimension~$x$, described by a wave function~$\phi(x,t)$ that depends
on time~$t$, is given by
\begin{eqnarray}
\left[ - \frac{\hbar^2}{2 M} \frac {\partial^2 }{\partial  x^2  }
- i
\hbar \,{\frac {\partial }{\partial t}} \right]  \phi \left( x,t
 \right)
 =0   \label{eq_FreeSchroedingerEq}
\end{eqnarray}
where $h$ is Planck's constant in $\hbar = h / (2 \pi)$.

Since the wave functions for free particles and those subjected to
harmonic potentials factorize with respect to their spatial
coordinates, we will only discuss the one-dimensional case, rather
than two or three dimensions.

The coordinate transformations~\cite{Takagi_PIT90,Ole_AJP05} from free
to trapped system
\begin{eqnarray}
x(\xi,\tau) = \frac{\xi \sqrt{b \; \omega}}{ \cos( \omega \tau ) }
\quad \mbox{and} \quad t(\tau) = b \tan( \omega \tau ) 
\label{eq_t_map}
\end{eqnarray}
conform with classical expectations, compare
section~\ref{sec_Classical}. Here $b$ is a free parameter to be fixed
below for the case of the modelling of instantaneous environmental
changes. Applied to the mapping of solutions $\phi(x,t)$ of
Eq.~(\ref{eq_FreeSchroedingerEq}), Eqs.~(\ref{eq_t_map}) yield the
wavefunction of a harmonically trapped particle
\begin{eqnarray}
\psi(\xi,\tau) & = & \frac{\phi(x(\xi,\tau), t(\tau)) }{
f(x(\xi,\tau),t(\tau);b)} , \label{eq_func_map}
\end{eqnarray}
which contains the form factor
\begin{eqnarray} 
f(x,t;b) & = & \frac{\exp \left( { \frac{i M
t x^2}{2 \hbar (t^2+b^2)}} \right) }{ (1+t^2/b^2)^{1/4}} ,
\label{eq_wavefactor}
\end{eqnarray}
and solves \schr's equation of a harmonic oscillator with spring
constant~$k$
\begin{eqnarray}
\left[ -  \frac{\hbar^2}{2 M} {\frac {\partial^2 }{\partial  \xi^2  }} 
- i \hbar \,{\frac {\partial }{\partial \tau}}  
+ \frac{ k}{2} \xi^2 \right]  \psi \left(
\xi,\tau \right) =0 .   \label{eq_HOSC_SchroedingerEq}
\end{eqnarray}
This is straightforward (and tedious) to check by direct substitution.

Transformation~(\ref{eq_t_map}) maps $t\in (-\infty, \infty )$ onto
$\tau \in (-\pi/2,\pi/2)$ only, but the periodicity arising through
the use of trigonometric functions represents the oscillator's motion
for all times~$\tau$.

It is noteworthy that the discontinuities of the
transformation~(\ref{eq_t_map}) map from a \schr equation with a
continuous to another with a discrete energy spectrum. Time and energy
being canonical conjugates, this conspires to give us the discrete
energy spectrum of the harmonic oscillator.

We note that the term $(1+t^2/b^2)^{-1/4}$ occurs in the form
factor~(\ref{eq_wavefactor}) to preserve the normalization of the
mapped wave function. Its numerator can be correctly guessed when one
attempts to cancel terms containing the factor $\xi \frac{\partial
  \phi}{\partial \xi}$. Otherwise such terms arise in
Eq.~(\ref{eq_HOSC_SchroedingerEq}); compare the discussion
following Eq.~(\ref{eq_mixed_general}) in
section~\ref{sec_generalized} below.

\subsection*{Inverse Map (Trapped to Free)}

The coordinate transformations inverse to~(\ref{eq_t_map}) are
\begin{eqnarray}
  \xi & = & \frac{x}{\sqrt{b \; \omega}\sqrt{ \left( 1+ t^2/b^2 
\right) }} 
\label{eq_xi_inv_map}
\\
\mbox{and} \qquad
\tau & = & \frac{1}{\omega} \arctan\left( \frac{t}{b} \right) \; ,
\label{eq_tau_inv_map}
\end{eqnarray}
and go together with the wave function multiplication~$ \phi = \psi f
$, inverting Eq.~(\ref{eq_func_map}) and thus mapping from
Eq.~(\ref{eq_HOSC_SchroedingerEq}) to
Eq.~(\ref{eq_FreeSchroedingerEq}).

\section{Instantaneous Changes of the Environment }\label{sec_DetailsMap}

\begin{figure}[t]
\centering
\includegraphics[width=0.48\textwidth,height=0.3\textwidth]{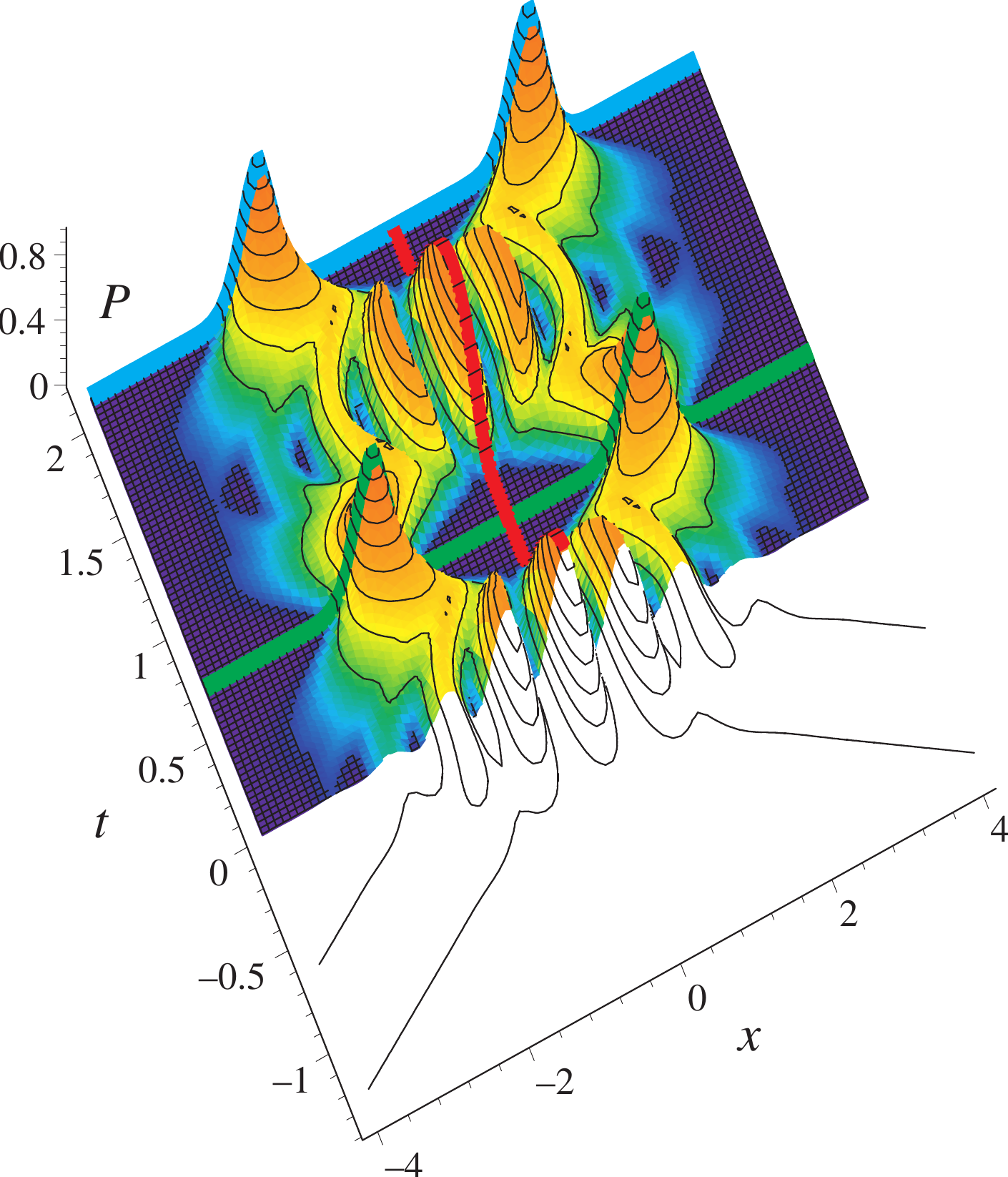}
\caption{Probability density of a free quantum particle
  in an equal superposition of two Gaussian states $P(x,t)\sim
  |\phi_0(x,t;0,p_0)+\phi_0(x,t;0,-p_0)|^2$ as described by
  Eq.~(\ref{eq_phi_0_x_t}), with $\hbar = 1$, $M = 1$, $\sigma_0 =
  3/2$, and opposing momenta~$p_0 = 4$. At time~$t=\tau=0$ the
  particle instantaneously gets trapped by a potential with spring constant
  $k=5$. The motion of the two half waves leads to interference near
  the origin of the potential (centered on red line at
  $x_0=\xi_0=0$). This plot illustrates that the trapped particle ends
  up in a superposition of two squeezed coherent
  states~\cite{Schleich_Book}. The squeezing (narrowing of peaks) is
  evident at times $T/4$ and $3 T/4$ after the transition time (green
  and blue cross lines; $T=2 \pi /\sqrt{5}\approx
  2.81$).} \label{fig_free_trapped_particle}
\end{figure}

If one wants to model instantaneous changes happening to the
environment of a quantum particle one generally has to decompose the
wave functions of the system at the time of the transition into the
superposition of eigenfunctions of the `new' system and then determine
their time evolution in the new system. This can involve sums over
infinitely many eigenstates of the new system making exact results
hard to obtain, often forcing us to truncate expressions.

In contrast, for maps between systems with potentials up to second
order in $\xi$, the approach discussed here can model instantaneous
transitions of the potential with ease. For several successive instantaneous
changes one ends up with the following sequence: determine the initial
state, then its time evolution in the free case, map onto the trapped
case (if that is to be modelled), determine its final state at the
next transition point in time~$t_0$, this serves as the initial state
of the next free propagation-step, etc.

It is desirable to shift the origin of time to each respective
transition point $t_0=0$ and it is necessary to choose the coordinate
stretching factor $b$ in such a way that the wave function is
momentarily unchanged at the transition point. Without loss of
generality the specific choices
\begin{eqnarray}
\tau_0 & = & \tau(t_0) = \tau(0) = 0 \label{eq_t_natural}
\\
\mbox{and} \qquad
b & = & 1/\omega = \sqrt{M/k}  \label{eq_b_natural}
\end{eqnarray}
yield such a smooth mapping since, according to Eq.~(\ref{eq_t_map}),
in this case
\begin{eqnarray}
x(t_0)=\xi(\tau_0) .
\label{eq_x_xi_smooth}
\end{eqnarray}
Eq.~(\ref{eq_x_xi_smooth}) assures that the two branches, say
$\phi(x,t)$ for $t<t_0=0$ and $\psi(\xi,\tau)$ for \mbox{$\tau
  >\tau_0=0$}, can be glued together continuously
\mbox{$\phi(x,0)=\psi(\xi,0)$}, since additionally 
for the `smooth' choice $b=1/\omega$ the form factor
$f(x,t_0;\frac{1}{\omega})=1$.


Note that the mapping implies that the \emph{entire environment} of
the particle changes instantaneously, irrespective of the particle's
quantum state. The energy of the particle is not conserved: mapping
from Eq.~(\ref{eq_FreeSchroedingerEq}) to
Eq.~(\ref{eq_HOSC_SchroedingerEq}) increases the particle's total
energy by the potential energy $\langle V \rangle = \int dx \;
\overline{\psi(\xi,t_0)} \frac{k}{2} \xi^2 \psi(\xi,t_0)$ (the overbar
denotes complex conjugation), the reverse map correspondingly reduces
it. This is quite different to the textbook scenario where a particle
approaches a region in space where the potential, at some
location~$x_0$, steps abruptly to a different potential. In that case
the particle's energy remains unchanged but its wave function changes
abruptly, including being reflected at the step.

To illustrate the application of the mapping to a instantaneous change of the
environment we use the well-known textbook example of a freely
evolving Gaussian wave-packet with initial position spread~$\sigma_0$
\begin{eqnarray}
\phi_0(x,t;x_0,p_0)
& = & \frac {1 }{\sqrt{\sqrt{\pi} \sigma(t)}}
\exp{\left[ {{-i M v_0 \frac {(x-x_0)\sigma_0 }{
 \hbar \sigma(t)}}}  \right]}
\nonumber \\
& \times & \exp{\left[   {-\frac { (x-x_0)^2}{2
\sigma_0 \sigma(t)}} - i {\frac {M{v_0}^2 }{2} \frac{t\sigma_0}{ \hbar
\sigma(t)}} \right]} , \quad {}^{} \label{eq_phi_0_x_t} \\
\mbox{where} \quad \sigma(t) & = & \sigma_0+i \frac{t\hbar}{\sigma_0
M} . \label{eq_sigma_t}
\end{eqnarray}
Here $x_0$ parameterizes spatial and $p_0 = M v_0 $ momentum
displacement of the wave functions. If either of these two quantities
is non-zero the mapping onto a harmonically trapped state results in a
state with oscillating center-of-mass. In general, although it is a
wave function of Gaussian shape, i.e. a coherent state, this freely
evolving wave packet will not `fit' the width of the harmonic
potential and therefore be `squeezed'. In short, the state of
Eq.~(\ref{eq_phi_0_x_t}) trapped in a harmonic potential becomes a
squeezed coherent state~\cite{Schleich_Book}, see cross lines in
Fig.~\ref{fig_free_trapped_particle}.

\section{Composed Maps (Trapped to Trapped)}\label{sec_compose}

The composed coordinate transformations from an initial harmonic
trapping potential with spring constant~$k$ and wave function $\psi
(\xi,\tau;k)$, via the free particle-case~$\phi$, to a final harmonic
potential with spring constant~$K$ and wave function
$\Psi(\xi,\tau;K)$ is given by
\begin{eqnarray}
\Xi & = &
\frac{\xi}{\sqrt{\cos^2( \Omega \tau) +\frac{k}{K}\sin^2(\Omega \tau)}}
\; , \label{eq_xi_Xi_map}
\\
T & = & \frac{1}{\omega} \arctan \left( \sqrt{\frac{k}{K}} \tan\left(
\Omega \tau \right) \right) ,
\qquad \label{eq_tau_Tau_map}
\end{eqnarray}
\begin{eqnarray}
\mbox{and} \quad \Psi(\xi,\tau;K)  & = & \psi(\Xi(\xi,\tau),T(\tau);k)
 \label{eq_func_map_hosc_HOSC}
 \\ 
 & \times & \frac{f(x(\xi,\tau),t(\tau);\frac{1}{\omega}) }{
   f(x(\xi,\tau),t(\tau);\frac{1}{\Omega})} . \quad \nonumber 
\end{eqnarray}

Here $\Omega=\sqrt{K/M}$, and~$\Psi(\xi,\tau;K)$ solves \schr
Eq.~(\ref{eq_HOSC_SchroedingerEq}) with $k$ substituted by~$K$.

It can be checked that the inverse of the coordinate
transformations~(\ref{eq_xi_Xi_map}) and~(\ref{eq_tau_Tau_map}) are
given by the same functional expressions with the quantities
pertaining to one potential swapped with those of the other ($\psi
\leftrightarrow \Psi$, $k \leftrightarrow K$ and $\omega
\leftrightarrow \Omega$).

\section{Laterally shifted Traps}\label{sec_LatShift}

\begin{figure}[b]
\centering
\includegraphics[width=0.48\textwidth,height=0.3\textwidth]{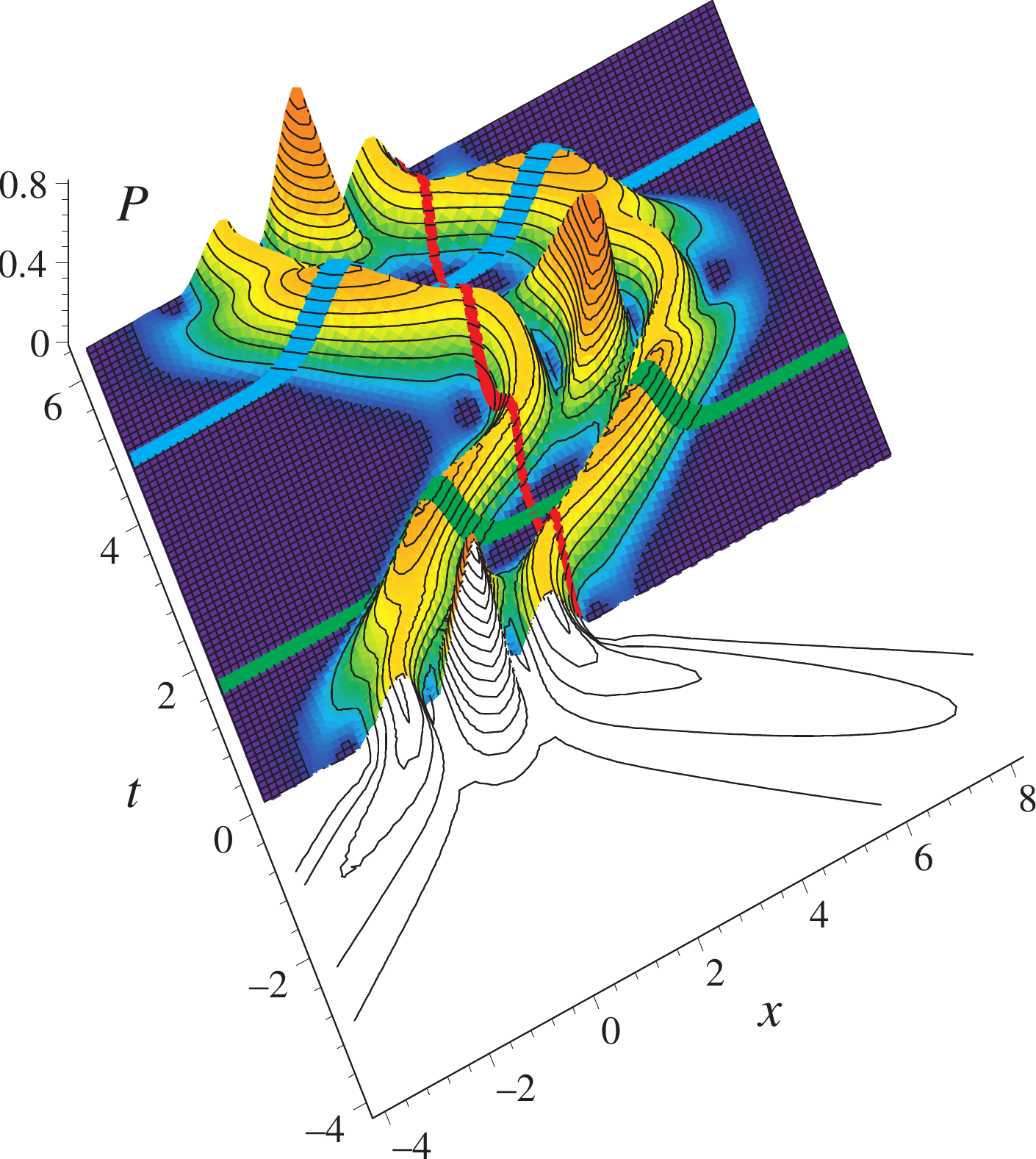}
\caption{Same as Fig.~\ref{fig_free_trapped_particle}
  above with the center of the trapping potential, indicated by the
  solid red line, laterally shifted to $\xi_0=2$. Here~$p_0=2$ and
  $k=1$ ($T=2 \pi \approx 6.28$).}
  \label{fig_free_trapped_displaced_particle} 
\end{figure}

The coordinate transformation~(\ref{eq_t_map}) can also incorporate a
lateral shift of the trapping potential. Direct substitution shows
that the mapping
\begin{eqnarray}
x(\xi-\xi_0,\tau) = \frac{(\xi - \xi_0) }{ \cos( \tau \omega ) }
\label{eq_x_mapshift}
\end{eqnarray}
applied to Eqns.~(\ref{eq_func_map}) and~(\ref{eq_wavefactor}) yields
Schr\"odinger's Eq.~(\ref{eq_HOSC_SchroedingerEq}) for the harmonic
oscillator with a shifted trapping potential: $k\xi^2 \mapsto
k(\xi-\xi_0)^2$.  This observation allows us to model instantaneous
transition of a particle from the free state or a trap centered at
some initial position to another trap with different stiffness and
center position, compare Fig.~\ref{fig_free_trapped_particle} with
Fig.~\ref{fig_free_trapped_displaced_particle} below. In order to
implement this map we use Eq.~(\ref{eq_x_mapshift}) to map the
position shift over into the new set of coordinates of the displaced
trapping potential and subsequently `undo' this shift by
substracting~$\xi_0$ from the argument of the wave function in order
to make sure that outgoing and incoming waves match up at the
transition time; compare discussion following
Eq.~(\ref{eq_b_natural}).  Note that this latter compensation only
applies to the wave function that is to be mapped, not the dividing
factor~$f(x,t;\frac{1}{\omega})$: Eq.~(\ref{eq_func_map}) thus becomes

\begin{eqnarray}
  \psi(\xi,\tau) & = & \frac{\phi(x(\xi-\xi_0,\tau) 
+ \xi_0,t(\tau)) }{f(x(\xi-\xi_0,\tau),t(\tau);\frac{1}{\omega})} .
\label{eq_func_map_shifted}
\end{eqnarray}

The same applies to equations~(\ref{eq_xi_Xi_map})
and~(\ref{eq_func_map_hosc_HOSC}).

\section{Inverted `Oscillators'}\label{sec_Inv_Osc}

In his foundational work on inverted `oscillators'
Barton~\cite{Barton_AnnP86} used that a regular harmonic oscillator
can be converted into an inverted oscillator using the formal
`complexification' $\omega \mapsto i \omega$. It was later realized by
Yuce \emph{et al.}~\cite{Yuce_PhysScrpt06} that a coordinate
transformation like~(\ref{eq_t_map}) allows for a map from free system
to inverted harmonic potentials. Using Barton's substitution and the
smooth choice~($b=\frac{1}{\omega}$) Eqs.~(\ref{eq_t_map}) become
\begin{eqnarray}
x(\xi,\tau) = \frac{\xi }{ \cosh( \tau \omega ) }
\quad \mbox{and} \quad t(\tau) =  \frac{\tanh( \tau \omega )}{|\omega|} \; .
\label{eq_map_free_inverted}
\end{eqnarray}
Wave function map~(\ref{eq_func_map}) and form
factor~(\ref{eq_wavefactor}) remain unchanged. This yields the
harmonic oscillator \schr equation~(\ref{eq_HOSC_SchroedingerEq}) with
spring constant~$ k\mapsto -k$ and can include the lateral
shift~(\ref{eq_x_mapshift}) as well.

\begin{figure}[t]
\centering
\includegraphics[width=0.48\textwidth,height=0.3\textwidth]{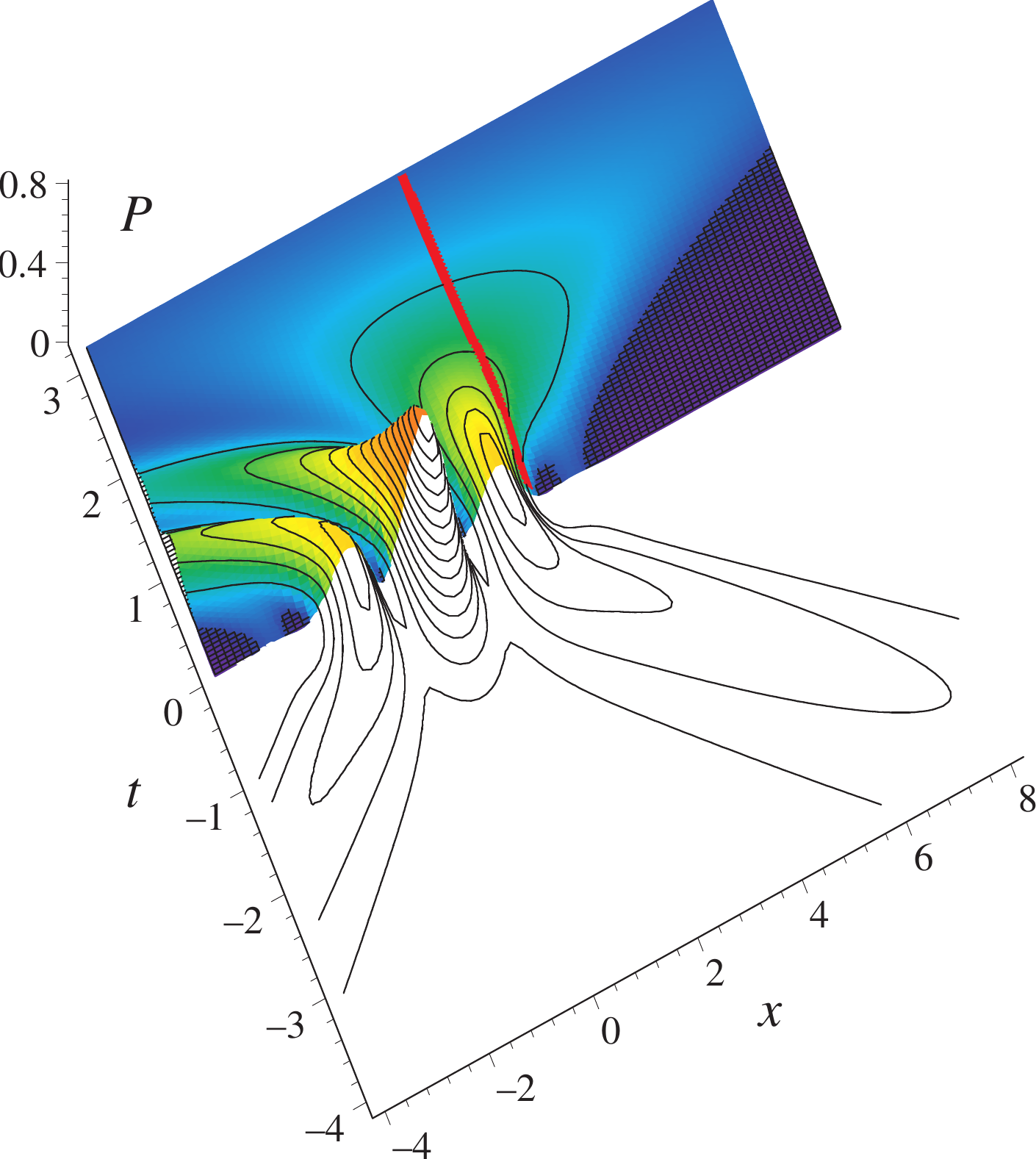}
\caption{Same as
  Fig.~\ref{fig_free_trapped_displaced_particle} above except for the
  potential being inverted: $k=-1$.}
  \label{fig_inverted_Osc_displaced_particle} 
\end{figure}

This case is illustrated in Fig.~\ref{fig_inverted_Osc_displaced_particle}.

\section{From Free to Freely Falling Particle}\label{sec_freeFall}

In order to provide further context for the transformations introduced
here, we now consider the mapping from a free to a freely falling
quantum particle. This simple case is readily treated with the
approach advertised here. As before, the free particle with wave
function $\phi(x,t)$ fulfils Eq.~(\ref{eq_FreeSchroedingerEq}), we
substitute
\begin{eqnarray}
x(\xi,\tau) = \xi + \frac{a \tau^2}{2}
\quad \mbox{and} \quad t(\tau) = \tau \; .
\label{eq_map_freefall}
\end{eqnarray}
Applied to the mapping of solutions of
Eq.~(\ref{eq_FreeSchroedingerEq})
\begin{eqnarray}
\psi(\xi,\tau) & = & \frac{\phi(x(\xi,\tau), t(\tau)) }{
g(x(\xi,\tau),t(\tau);a)} , \label{eq_func_map_free_case}
\\
\mbox{with} \nonumber
\\
 g(x,t;a) & = & 
\exp\left(i \frac{M a t}{\hbar}( x + \frac{a}{2} t^2)+ i \frac{M a^2}{6 \hbar}  t^3 \right) ,
\label{eq_wavefactor_freefall}
\end{eqnarray}
this yields solutions~$\psi(\xi,\tau)$ for the \schr equation
\begin{eqnarray}
\left[ -  \frac{\hbar^2}{2 M} {\frac {\partial^2 }{\partial  \xi^2  }} 
- i \hbar \,{\frac {\partial }{\partial \tau}}  
+ M a \xi \right]  \psi \left(
\xi,\tau \right) =0 \quad{}^{},   \label{eq_FF_SchroedingerEq}
\end{eqnarray}
where $a$ is the acceleration of the freely falling particle.

\section{Can this Method be generalized?\label{sec_generalized}}

Our investigations raise the question whether the method described
here can be generalized to other potentials, perhaps to
\emph{anharmonic} potentials such as the solvable cases discussed in
references~\cite{Gendenshtein_JETP83} and~\cite{Cooper_JPA89}?

One might think recent work by Costa Filho \emph{et
  al.}~\cite{CostaFilho_EPL13} provides an example of just such a
mapping from the harmonic case to that of the (anharmonic) Morse
potential. But it turns out that the coordinate transformation
employed there does not preserve the standard commutation relations of
quantum physics (or Heisenberg's uncertainty
principle~\cite{CostaFilho_EPL13}) rendering the transformation
unphysical.

The phase space time evolution of a classical harmonic oscillator is
known to amount to a rigid rotation because the oscillation frequency
is independent of the state, the amplitude, of the oscillator. For
anharmonic potentials this state independence does not apply which is
why they are very different to harmonic potentials.  Recent
investigations of Wigner flow in quantum phase space~\cite{Ole_PRL13}
have established that anharmonic potentials show qualitatively very
different quantum phase space flow behaviour from the rigid rotation
that is found in the harmonic case. It therefore appears unlikely that
a simple coordinate transformation could map from one to the other
thus extending the equivalence to anharmonic potentials.

We will now show that, indeed, no mapping from the studied three
equivalent cases (free particle, freely falling particle and particle
in harmonic potential) to cases with anharmonic potentials exists. We
first investigate the constraints on coordinate transformations of the
general form
\begin{eqnarray}
x(\xi,\tau) = X(\xi, \tau)
\quad \mbox{and} \quad t(\xi,\tau) = T (\xi,\tau) \; .
\label{eq_generalized_map}
\end{eqnarray}
The transformations Eq.~(\ref{eq_t_map}), Eq.~(\ref{eq_tau_inv_map})
and Eq.~(\ref{eq_map_freefall}) introduced so far have in common that
the mapped time~$\tau$ only depends on the original time $T(\xi,\tau)
= T(\tau)$. If $T$ depended on $\xi$ then $\partial T /\partial \xi
\neq 0$ which in turn would imply that the differential operator
$\partial^2/\partial x^2$, upon transformation to the new coordinates,
`spills over' into the mapped time giving rise to a
term~$\partial^2/\partial \tau^2$ which cannot be cancelled by other
terms. This can be shown by applying the chain rule or studying the
Hessian of the transformation. Since second order time derivatives
have no place in \schr's equation we have established that the
transformation for time never depends on the spatial coordinate.

With $X' = \partial X(\xi,\tau) /\partial \xi$, $X'' = \partial^2
X(\xi,\tau) /\partial \xi^2$ and $\dot T = \partial T(\tau)/\partial
\tau$, application of the chain rule yields
\begin{eqnarray}
\frac{\partial}{\partial t}
& = & \frac{1}{\dot T} \frac{\partial}{\partial \tau}
\label{eq_chain_rule_X}
\\
\mbox{and} \quad 
\frac{\partial^2}{\partial x^2} & = & 
\frac{1}{{X'}^2} \frac{\partial^2}{\partial \xi^2}
-\frac{X''}{{X'}^3} \frac{\partial}{\partial \xi}
+ {O}(\frac{\partial}{\partial \tau}) \; ,
\label{eq_chain_rule_T}
\end{eqnarray}
where ${O}(\frac{\partial}{\partial \tau})$ refers to terms containing up to first order
derivatives in $\tau$. Obviously, the transformation of partial derivatives `incurs'
coefficient functions~$C(\xi,\tau)$.  The first coefficients in
Eqs.~(\ref{eq_chain_rule_X}) and~(\ref{eq_chain_rule_T}) need to be equal, that is
\begin{eqnarray}
C(\xi,\tau) = \frac{1}{\dot T} =  \frac{1}{{X'}^2}\; .
\label{eq_chain_rule_Equality}
\end{eqnarray}
Only this allows us to extract the common factor~$C$ guaranteeing that
the mapping preserves the structure of \schr's equation for the sum of
total and kinetic energy
\begin{eqnarray}
  - \frac{\hbar^2}{2M} \frac{\partial^2}{\partial x^2} - i \hbar \frac{\partial}{\partial t}
  \mapsto C \left[ - \frac{\hbar^2}{2M} 
    \frac{\partial^2}{\partial \xi^2} - 
    i \hbar \frac{\partial}{\partial \tau} \right] 
  + \cal{V}; \qquad
\label{schroedinger_structure_preserved}
\end{eqnarray}
here $\cal{V}$ refers to all other terms. With
Eq.~(\ref{eq_chain_rule_Equality}) we have established that
\begin{eqnarray}
T  = \int d \tau \; {X'(\xi,\tau)}^2\; ,
\label{eq_T_int_Xp2}
\end{eqnarray}
at the same time we know that $T$ must not depend on the spatial
coordinate $\xi$, therefore $X$ is at most \mbox{linear in $\xi$}
\begin{eqnarray}
X(\xi,\tau) = A(\tau) \xi + B(\tau) \; ,
\label{eq_X_lin_xi}
\end{eqnarray}
and thus
\begin{eqnarray}
T(\tau) = \int_{\tau_0}^{\tau} d \tilde \tau A( \tilde \tau)^2 \; .
\label{eq_T_A2}
\end{eqnarray}
Here $A$ and $B$ are, at this stage, arbitrary functions of $\tau$
such that all transformations are invertible. We write the form
factor~$f$ as
\begin{eqnarray}
  f(\xi,\tau) = \frac{\exp[i \epsilon(x(\xi,\tau),t(\xi,\tau))]}{\sqrt{N(\tau)}} \; ,
\label{eq_f_general}
\end{eqnarray}
where the normalization $N(\tau) = A(\tau)$ takes account of the
stretching of the coordinate system and $\epsilon$ is real.  This,
respectively, enforces normalization of $\psi$ and provides a general
unitary transformation for it. In the mapped \schr equation it
generates a mixed term of the transformed differentials (hidden in the
symbol~$\cal V$ in Eq.~(\ref{schroedinger_structure_preserved}))
proportional to the momentum operator $\frac{\hbar}{i}\frac{\partial
}{\partial \xi}$
\begin{eqnarray}
\left(\frac{\hbar}{M} \frac{\partial \epsilon(\xi,\tau)}{\partial \xi} 
- \frac{1}{A(\tau)} \frac{\partial [\xi A(\tau)+B(\tau)]}{\partial \tau} \right) 
\frac{\hbar}{i} \frac{\partial \psi(\xi,\tau)}{\partial \xi} . \qquad
\label{eq_mixed_general}
\end{eqnarray}
Such a term has no place in \schr's equation, since we do not want to model fields
coupling to the particle's momentum. The function in round brackets in
Eq.~(\ref{eq_mixed_general}) has to be zero; its integration with respect to $\xi$ yields
\begin{eqnarray}
  \epsilon(\xi,\tau) 
  = \frac{M}{\hbar A(\tau)} \frac{\partial}{\partial \tau} 
  \left[\frac{\xi^2}{2} A(\tau)+\xi B(\tau) + \Xi_0(\tau)\right]  . \qquad
\label{eq_condition_on_epsilon}
\end{eqnarray}
Having rid ourselves of the momentum term~(\ref{eq_mixed_general}) by
insertion of~(\ref{eq_condition_on_epsilon}) into~(\ref{eq_f_general})
we turn to the remainder~$\cal V$ in
Eq.~(\ref{schroedinger_structure_preserved}). It now represents the
time-dependent quadratic potential
\begin{eqnarray} {\cal V}(\tau) & = & \frac{\xi^2}{2} \frac{M}{A(\tau)^2}
  \left[ A \left( \tau \right) \left( {\frac {d^{2}}{d{\tau}^{2}}}A \left(
      \tau \right) \right) -2 \left({\frac {d}{d\tau}}A \left( \tau \right)
    \right) ^{2} \right]
 \nonumber \\
  & + & \xi M \left[ {\frac {{\frac {d^{2}}{d{\tau}^{2}}}B \left( \tau
        \right) }{A \left( \tau \right) }} -2\,{\frac { \left( {\frac
            {d}{d\tau}}B \left( \tau \right) \right) \left( {\frac
          {d}{d\tau}}A \left( \tau \right) \right) }{ A \left( \tau
          \right)^{2}}} \right]
 \nonumber \\
  & + & M \left[{\frac {d}{d\tau}}\Xi_{{0}} \left( \tau \right)
    -{\frac { \left( {\frac {d}{d\tau}}B \left( \tau \right)
        \right) ^{2}}{2  A \left( \tau \right) ^{2}}}
  \right] . \qquad
\label{eq_general_V}
\end{eqnarray}
Evidently only quadratic potentials with time-dependent
frequency~\cite{Takagi_PIT90} can be generated with our approach, the
method cannot be generalized to anharmonic potentials!

Through changes of $A$, $B$ and $\Xi_0$ some freedom of choice is
available for time-dependent quadratic potentials. For time-dependent
quadratic potentials Lewis established a constant of motion in
1967~\cite{Lewis_PRL67,Lewis_JMP68}, Ray applied it in
1982~\cite{Ray_PRA82}, as did Agarwal~\cite{Agarwal_PRL91} and many
others. A recent overview is given in Lohe's 2009
work~\cite{Lohe_JPA09}.

To recap our previous examples: Eq.~(\ref{eq_FF_SchroedingerEq}) is
reproduced by $\Xi_0(\tau)= \frac{1}{2} \int d \tau \left( \partial
  B(\tau)/\partial \tau \right)^2 $ with $A = 1,$ and $B(\tau) =
\frac{a}{2} \tau^2$. $\Xi_0(\tau)=0$, $A(\tau)=\exp(i\omega\tau)$ and
$B=0$ yields the \schr equation of the harmonic
oscillator~(\ref{eq_HOSC_SchroedingerEq}), whereas its
`complexification' $A(\tau)=\exp(\omega\tau)$ results in the \schr
equation for the inverse oscillator.

\begin{figure}[t]
\begin{center}
  \begin{minipage}[t]{0.99\linewidth}
    \includegraphics[width=0.99\linewidth,height=0.69\linewidth]{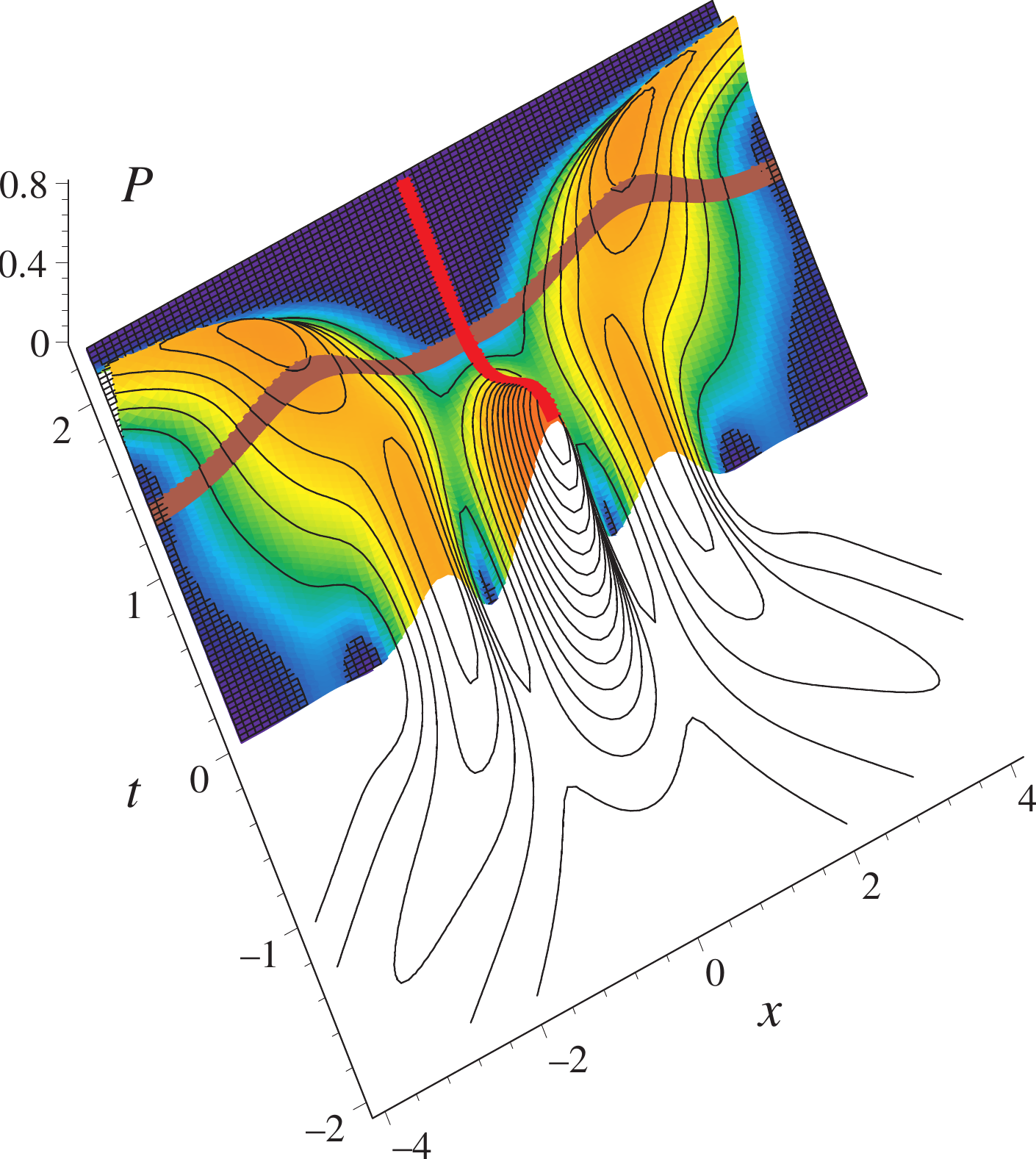}
\put(-65,101){\rotatebox{0}{\fcolorbox{black}{white}{\includegraphics[width=0.25\linewidth,height=0.18\linewidth]{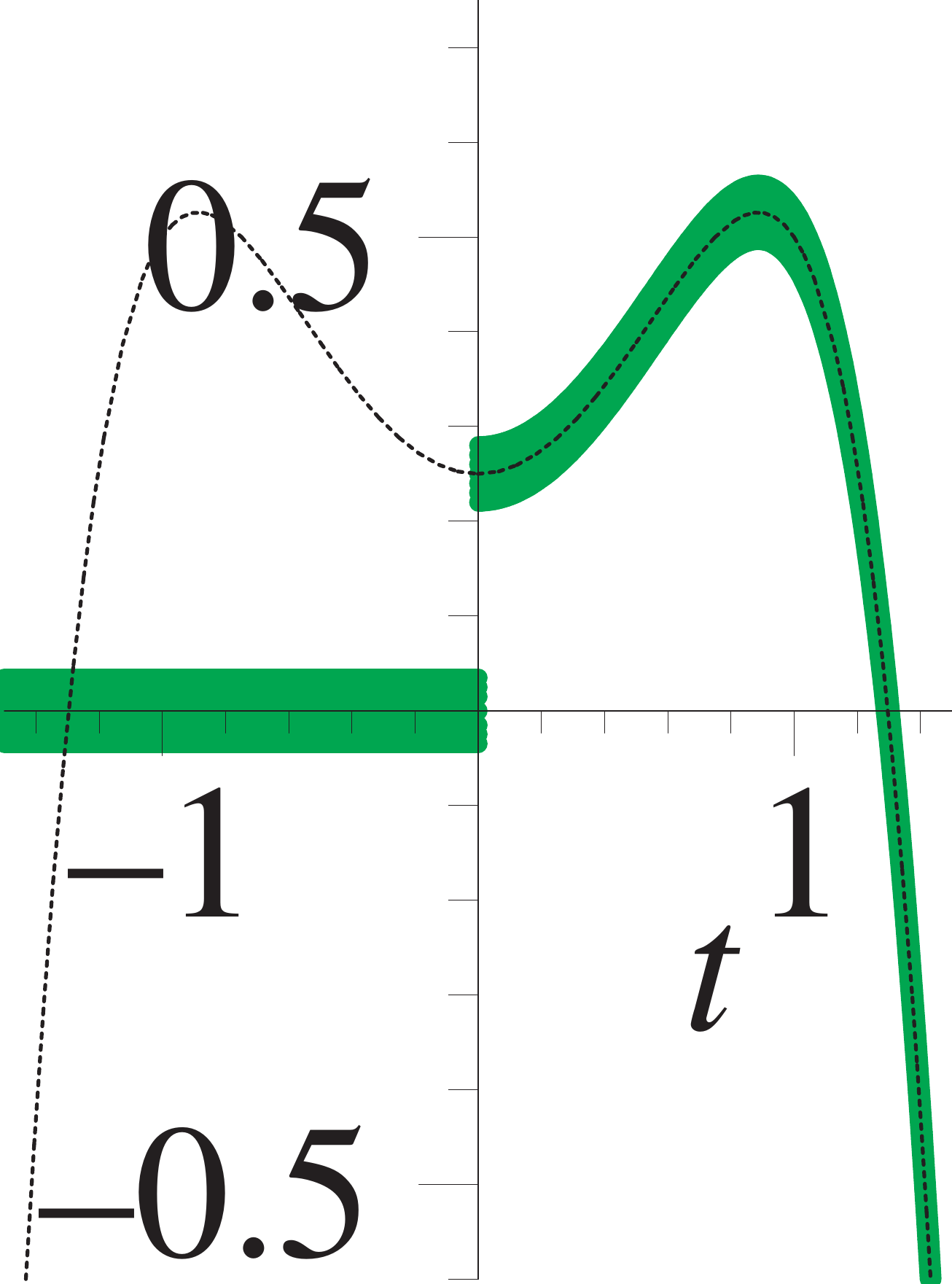}}}}
\end{minipage}
\end{center}
\caption{The time-dependence of the potential's
  quadratic term ${\cal V}_2/\xi^2$, which can be derived from $A_1$
  of Eq.~(\ref{eq_A1}), is shown as a thin line in the inset.  The
  thick green line in the inset describes a scenario where the
  potential is switched on at the transition time $\tau_0=0$, as is
  shown in the main figure. The free particle's wavefunction arrives
  centered with the minimum of the potential (thick red line) in a
  superposition just like in Fig.~\ref{fig_free_trapped_particle}
  (here~$p_0=2$).  The time-dependence of the potential gives rise to
  focusing of the wavefunction until, at time $\tau =1.35$ (thick
  brown line), the potential becomes
  repulsive. \label{fig_4_timedep_symmetric}}
\end{figure}

\begin{figure}[t]
\begin{center}
  \begin{minipage}[t]{0.99\linewidth}
    \includegraphics[width=0.99\linewidth,height=0.69\linewidth]{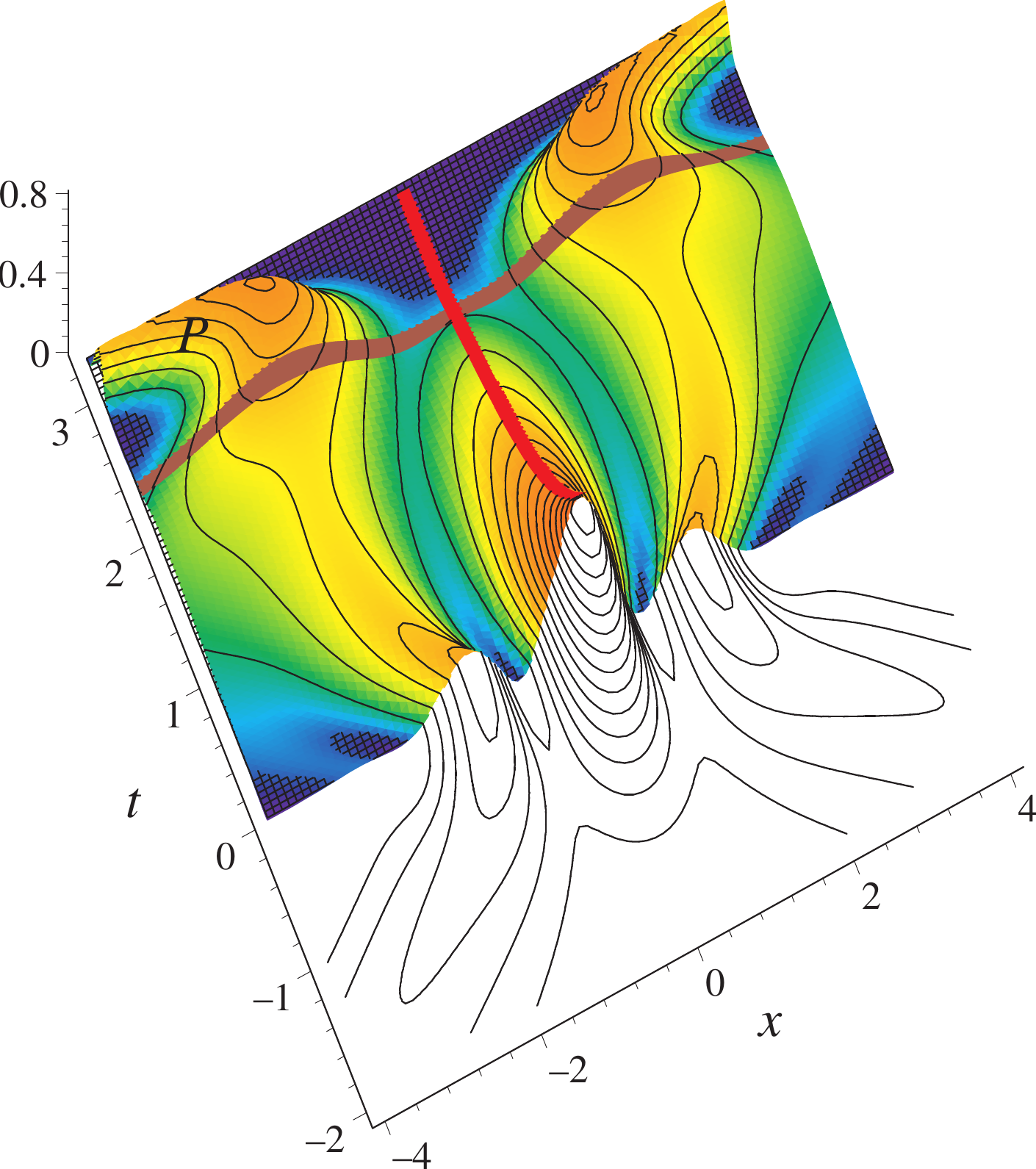}
\put(-65,101){\rotatebox{0}{\fcolorbox{black}{white}{\includegraphics[width=0.25\linewidth,height=0.18\linewidth]{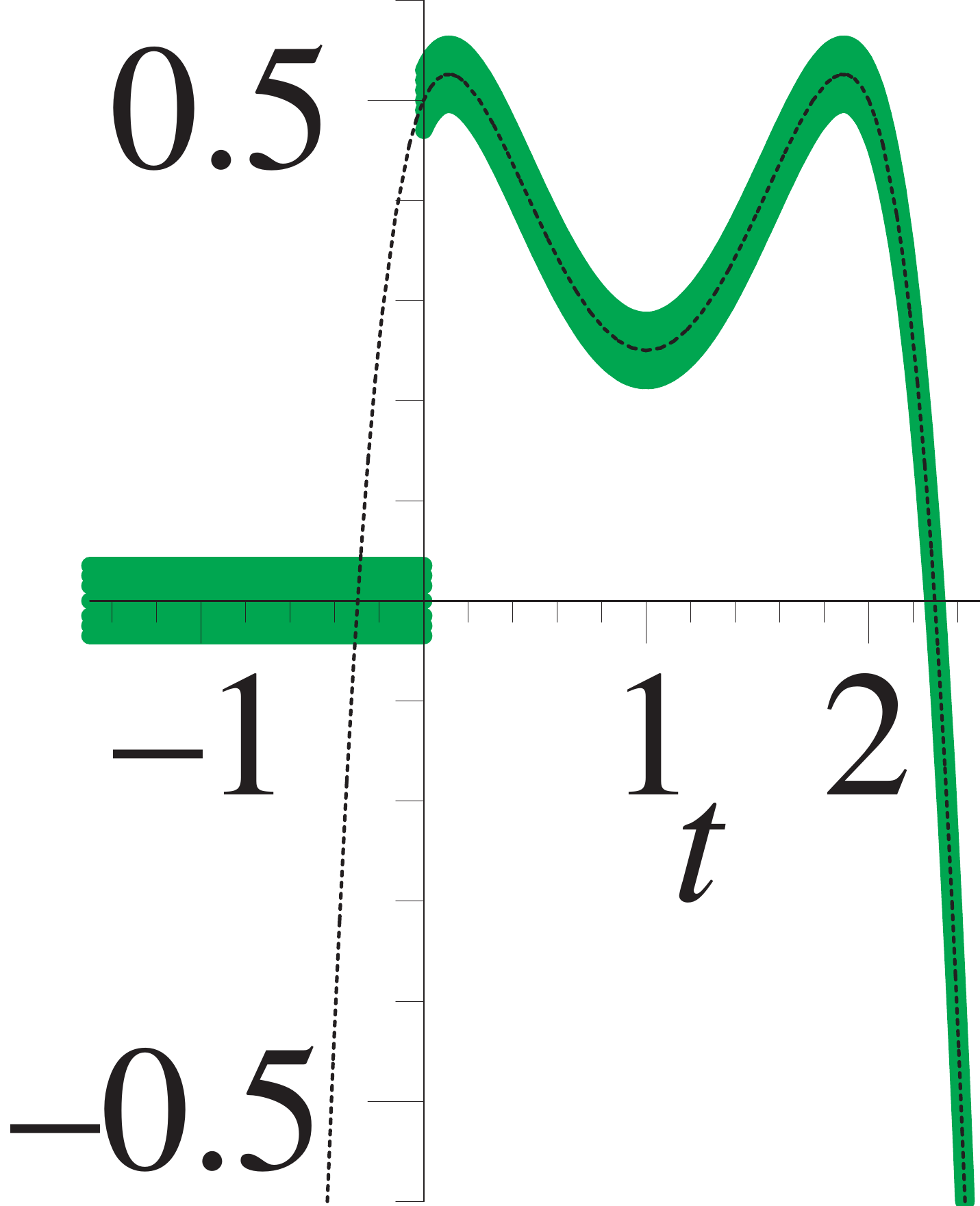}}}}
\end{minipage}
\end{center}
\caption{Essentially the same as
  Fig.~\ref{fig_4_timedep_symmetric}, here the time-dependence of the
  potential's quadratic term ${\cal V}_2/\xi^2$ is derived from $A_2$
  of Eq.~(\ref{eq_A2}). After the potential is switched on at
  $\tau_0=0$ it is attractive until at time $\tau = 2.3$ (thick brown
  line) it becomes repulsive. \label{fig_5_timedep_asymmetric}}
\end{figure}

\section{Time-dependent Harmonic Potentials\label{sec_timeDependent}}

We now turn to the general expression for the time-depen\-dent harmonic potentials, those
that conform with the form of~$\cal V(\tau)$ found in Eq.~(\ref{eq_general_V}). The
reference level of the energy can be chosen in any way we like because the
function~$\Xi_0(\tau)$ gives us complete freedom of choice. We will not investigate the
term linear in~$\xi$, we choose $B(\tau)=0$. The harmonic term in $\cal V$ has to conform
with the non-linear ordinary differential expression of $A(\tau)$ that forms its
coefficient function. This form was previously derived in 1996 by Bluman and Shtelen who
argue that it is `arbitrary'~\cite{Bluman_JPA96}. Unfortunately, this does not imply, that
it can be determined freely --as one pleases, see
Mostafazadeh~\cite{Mostafazadeh_JMP99}. Substituting $A(\tau)=\exp [\alpha(\tau)]$ the
quadratic term of $\cal V$ takes the alternative form
\begin{eqnarray} {\cal V}_Q(\tau) = \frac{\xi^2 M}{2} \left[{\frac
      {d^{2}\alpha}{d{\tau}^{2}}} - \left(\frac
      {d\alpha}{d{\tau}}\right)^2 \right] .
\label{eq_VQ}
\end{eqnarray}

Here, our main interest is in the application to instantaneous switching of
the potential; at the transition time $\tau=\tau_0$. In this case, for
either version of the form factor~(\ref{eq_f_general}), be it
expressed in terms of $\alpha$ or $A$, we find that the condition for
smooth coordinate linkage~(\ref{eq_x_xi_smooth}) requires that
\begin{eqnarray}
A(\tau_0) = 1 \mbox{ or } \alpha(\tau_0) = 0 \; .
\label{eq_A_tau0}
\end{eqnarray}

Two examples illustrate the findings of this section. Firstly, for Fig.~\ref{fig_4_timedep_symmetric}, we choose
\begin{eqnarray}
A_1(\tau) = \exp [\frac{1}{8} \tau^4 + \frac{1}{4} \tau^2 ] 
\label{eq_A1}
\end{eqnarray}
with switch-over time $\tau_0 = 0$ this conforms
with~(\ref{eq_A_tau0}). It yields a complicated expression for the
time-dependence of the harmonic potential with a simple graphical
representation, a positive double-hump that drops sharply to negative
values for large values of time, compare inset of
Fig.~\ref{fig_4_timedep_symmetric}.

Secondly, we shift $A_1(\tau - \frac{1}{2})$ and divide by~$A(0)$, to
conform with~(\ref{eq_A_tau0}), the resulting expression
\begin{eqnarray}
  A_2(\tau)& = &
\exp [\frac{1}{8} (\tau-1)^4 + \frac{1}{4} (\tau-1)^2 -\frac{3}{8} ] 
\label{eq_A2}
\end{eqnarray}
was employed to generate Fig.~\ref{fig_5_timedep_asymmetric}.

Note that ${\cal V}_Q$ is non-linear in $A$ or $\alpha$ which makes
even changing the strength of the potential difficult to achieve.
Unfortunately, at this stage it appears unlikely that one should be
able to succeed in tailoring the time dependence of the harmonic
potential to one's needs exactly and still be able to determine $A$ or
$\alpha$, by some kind of inversion of Eq.~(\ref{eq_VQ}). Instead, one
is therefore forced to use Eqs.~(\ref{eq_general_V}) or~(\ref{eq_VQ})
to determine $\cal V$ from $A$ or $\alpha$. At least inversion of
$A$ or $\alpha$ is not needed since all relevant
expressions~(\ref{eq_X_lin_xi}), (\ref{eq_T_A2})
and~(\ref{eq_condition_on_epsilon}) are given in terms of $A$ or
$\alpha$,~$B$ and~$\Xi_0$.

It is, of course, possible to glue together several (or many) slices of
maps that approximate a scenario one wants to model.

\section{Conclusions\label{sec_concl}}

General solutions of the free-particle \schr equation can be mapped
onto solutions of the \schr equation for the harmonic oscillator using
a simple coordinate transformation~(\ref{eq_t_map}) in conjunction
with a multiplication of the wave function by a suitable form
factor~(\ref{eq_wavefactor}). This map is invertible and a composition
of two such maps allows us to map from one harmonic oscillator to
another with a different, positive or negative, spring constant and
different center position. The simplicity of the approach described
here makes it a tool of choice for the description of the wave
function of a particle experiencing instantaneous transitions from a
free to a harmonically trapped state, the instantaneous release from a
harmonic trap~\cite{Ole_arXiv1109.1818} or the instantaneous change of
the stiffness and center position of its harmonic trapping potential.

Instantaneous transitions for particles that are free to ones that are
trapped, or vice versa, can experimentally be implemented in quantum
optics using optical forces exerted on cold atoms for
trapping~\cite{Chu_RMP98,Grimm_review99},
movement~\cite{Walther_PRL12,Bowler_PRL12}, release and dispersal
(using blue-detuned, repulsive trapping fields~\cite{Grimm_review99}).
Similarly, molecules~\cite{Shuman_NAT10,Manai_PRL12} and much larger
micro-mechanical quantum
systems~\cite{Kippenberg_SCI08,Marquardt_Physics.2.40} can be
manipulated using optical forces. In all these cases the extent of the
light field tends to be much greater than the size of the manipulated
quantum particle and, since light travels fast, changes to the light
field affect the particle's global environment immediately. Switching
the light on or off, moving the center of an optical trap or changing
the frequency or intensity of the used light is frequently well
described by trapping potentials whose spring constant or center
position are modified instantaneously \emph{``everywhere''}.  All such
transitions are easily described with the theory presented here.

The mapping introduced here is computationally more efficient than
state-projection or harmonic oscillator propagator techniques and
conceptually simpler than mapping techniques such as those used for
super\-symmetric
potentials~\cite{Gendenshtein_JETP83,Cooper_JPA89}. It helps with
symbolic calculations because it allows for the determination of the
time evolution of wave functions trapped in a harmonic potential using
the lower computational overheads of wave functions evolving in free
space~\cite{Ole_arXiv1109.1818}.

For numerical calculations the trapped case provides better control
than the free case since the particle is spatially confined. One can
determine the largest momentum components and thus work out the
maximal harmonic oscillation amplitude and smallest expected feature
size in space. In time, on account of the harmonic oscillator's
periodicity, at most half an oscillation period is
needed. Accordingly, a suitable static grid for numerical
calculations is easily devised.

 I would like to thank Georg Ritter, Dimitris Kakofengitis, and
Gary Bowman for valuable feedback on the manuscript. 

\section*{Note added in proofs} I am grateful to Julio Guerrero for pointing me to
references~\cite{FernandezGuasti_PRA03,Carinena_S08,Guerrero_JPA11} introducing similar
concepts as discussed here using abstract conservation laws.

\bibliography{FreeHOSC}

\end{document}